\documentclass[prl,twocolumn,superscriptaddress,showpacs,nobalancelastpage]{revtex4}

\usepackage{graphicx}
\begin{document}

\title{Scheme for teleportation of quantum states 
onto a mechanical resonator}

\author{Stefano Mancini}
\author{David Vitali} 
\author{Paolo Tombesi} 
\affiliation{INFM and Dipartimento di Fisica, 
Universit\`a di Camerino, I-62032 Camerino, Italy}

\date{\today}

\begin{abstract}
We propose an experimentally feasible scheme to teleport
an unkown quantum state onto the vibrational degree of 
freedom of a macroscopic mirror. 
The quantum channel between the two parties is 
established by exploiting radiation pressure
effects. 
\end{abstract}


\maketitle

Teleportation of an unknown quantum state is its immaterial
transport through a classical channel \cite{BEN93}, employing one of the 
most puzzling resources of Quantum
Mechanics: entanglement
\cite{EIN35}.
A variety of possible experimental schemes have been proposed and few of
them partially realized in the discrete variable case
involving the polarization state of single photons\cite{BOU97,BOS98,JEN02}.
A successful achievement has been then obtained in the continuous variable case
of an optical field \cite{FUR98}.
However, the tantalizing problem of
extending quantum teleportation at the macroscopic scale
still remains open.

Recently, in the perspective of demonstrating and manipulating
the quantum properties of bigger and bigger objects \cite{JUL01},
it has been shown \cite{PRL02} how it is possible
to entangle two massive macroscopic oscillators,
like movable mirrors, by using radiation pressure effects.
The creation of such an entanglement at the macroscopic level suggests
an avenue for achieving teleportation of a continuous variable state
of a radiation field onto the vibrational state of a mirror.

\begin{figure}
\includegraphics[width=3.2in]{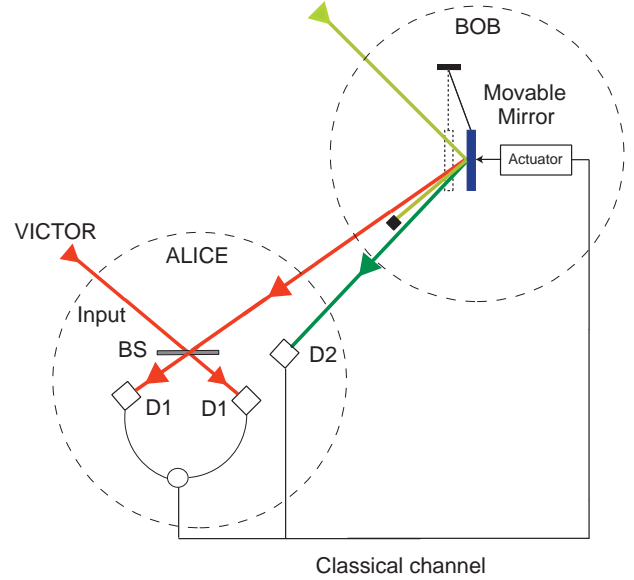}
\caption{
Schematic description of the system. 
A laser field at frequency
$\omega_{0}$ impinges on the mirror oscillating at frequency $\Omega$.
In the reflected field two sideband modes are excited at
frequencies $\omega_{1}=\omega_{0}-\Omega$ and
$\omega_{2}=\omega_{0}+\Omega$.
These two modes then reach Alice's station.
The mode at frequency $\omega_{2}$ is subjected to a heterodyne
measurement $D2$, while the mode at frequency $\omega_{1}$
is mixed in the 50-50 beam splitter BS with the unknown input given by Victor.
A Bell-like measurement $D1$ is then performed
on this combination and the result, combined with the
heterodyne one, is fed-forward to Bob.
Finally, he actuates the displacement
in the phase space of the moving mirror.
}\label{fig1}
\end{figure}

We consider the usual situation where an unknown
quantum state of a radiation field is prepared by a verifier
(Victor) and sent to an analyzing station (Alice).
Here we shall provide a protocol
which enables Alice to teleport the continuous variable
quantum state of the radiation
onto a collective vibrational degree of freedom
of a macroscopic, perfectly reflecting, mirror placed at a remote station
(Bob) (see Fig.~\ref{fig1}).
The mirror could also represent the cantilever of 
a micro-electro-mechanical system (MEMS) \cite{CLE98}.

For simplicity we consider only the motion and the elastic deformations
of the mirror taking place along the spatial direction $x$,
orthogonal to its reflecting surface. 
Then we consider an intense laser beam
impinging on the surface of the mirror, 
whose radiation pressure realizes an optomechanical coupling \cite{SAM95}.
In fact, the electromagnetic field exerts a force on the mirror 
proportional to its intensity and, at the same time,
it is phase-shifted by
the mirror displacement from the equilibrium position \cite{LAW95}.
In the limit of small mirror displacements, and in the interaction
picture with respect to the free Hamiltonian of the electromagnetic field
and the mirror displacement field ${\hat x({\bf r},t)}$
(${\bf r}$ is the coordinate on the mirror surface), one has the
following Hamiltonian
 \cite{PIN99}
\begin{eqnarray}
    {\hat H}&=&-\int\,d^{2}{\bf r}\,
    {\hat P}({\bf r},t){\hat x}({\bf r},t)\,,
    \label{eq:Hini}
\end{eqnarray}
where ${\hat P({\bf r},t)}$ is the radiation pressure force \cite{SAM95}.
All the continuum of electromagnetic modes
with positive longitudinal wave vector $q$, transverse
wave vector ${\bf k}$, and frequency $\omega=\sqrt{c^{2}(k^{2}+q^{2})}$
($c$ being the light speed in the vacuum)
contributes to the radiation pressure force.
The mirror displacement ${\hat x({\bf r},t)}$ is generally given by a 
superposition of many acoustic modes \cite{PIN99};
however, a single vibrational mode description can be adopted whenever 
detection is limited to a frequency bandwidth
including a single mechanical resonance. 
In particular, focused light beams are able to excite 
Gaussian acoustic modes, in which only a small portion of the mirror,
localized at its center, vibrates. These modes have a small  
waist $w$, a very large mechanical quality 
factor $Q$, and a small effective mass $M$ \cite{PIN99}. 
The simplest choice is to choose the fundamental Gaussian mode with 
frequency $\Omega$ and ladder operators ${\hat b}$, ${\hat b}^{\dag}$,
$[b,b^{\dagger}]=1$, so that
\begin{equation}
    {\hat x}({\bf r},t)=\sqrt{\frac{\hbar}{2M\Omega}}
    \left[{\hat b}e^{-i\Omega t}
    +{\hat b}^{\dag}e^{i\Omega t}\right]
    \exp(-r^{2}/w^{2})\,.
\end{equation}
The interaction Hamiltonian (\ref{eq:Hini}) assumes a simple form
if we average it over the time interval corresponding to the inverse 
of the spectral resolution of the optical detection apparatus at 
Alice station $\Delta \nu_{det}$, and if we choose $\Delta \nu_{det}
\ll \Omega$. This implies
neglecting all the terms oscillating faster than $\Omega$
(Rotating Wave Approximation, RWA). We also consider an intense 
driving laser polarized parallel to the mirror surface, 
with power $\wp$, bandwidth $\Delta \nu_{mode}$ centered around the 
carrier frequency $\omega_{0}$, impinging on the mirror 
with an angle of incidence $\phi_{0}$. 
If it is sufficiently intense, it can be treated as classical, 
and it is reflected undisturbed by the mirror. At the same time, it 
strongly couples the acoustic mode with the two back-scattered sideband  
modes at frequencies
$\omega_{1}=\omega_{0}-\Omega$ (with annihilation 
operator ${\hat a}_{1}$) and $\omega_{2}=\omega_{0}+\Omega$ 
(with annihilation operator ${\hat a}_{2}$). 
Thus, we end up with the three-mode, effective interaction Hamiltonian
\cite{LONG}
\begin{equation}
    {\hat H}_{eff}=-i\hbar \chi
    ({\hat a}_{1}{\hat b}-{\hat a}^{\dag}_{1}{\hat b}^{\dag})
    -i\hbar\theta({\hat a}_{2}{\hat b}^{\dag}-{\hat
    a}^{\dag}_{2}{\hat b})\,,
    \label{eq:Heff}
\end{equation}
where
$\chi=\cos\phi_{0} \sqrt{\wp\Delta\nu_{det}^{2}\omega_{1}/2M
\Delta\nu_{mode}c^{2}\Omega}$
and $\theta=\chi\sqrt{\omega_{2}/\omega_{1}}$.
The physical process described by this interaction Hamiltonian is 
very similar to a stimulated Brillouin scattering \cite{PER84}, 
even though, in the present case, the Stokes (${\hat a}_{1}$) and anti-Stokes 
(${\hat a}_{2}$) components are back-scattered by the acoustic waves at 
reflection, and the optomechanical coupling is provided by the 
radiation pressure and not by the dielectric properties of the mirror.

Eq.~(\ref{eq:Heff}) contains two interaction terms: the first one,
between modes ${\hat a}_{1}$ and ${\hat b}$,
is a parametric-type interaction
leading to squeezing in phase space \cite{QO94}, and it is
able to generate the EPR-like
entangled state which has been used in the continuous variable teleportation
experiment of Ref.~\cite{FUR98}. The
second interaction term, between modes ${\hat a}_{2}$ and ${\hat b}$,
is a beam-splitter-type
interaction \cite{QO94}, which may disturb the entanglement between
modes ${\hat a}_{1}$ and ${\hat b}$ generated by the first term.

The system dynamics can be easily studied through 
the (normally ordered) characteristic function $\Phi(\mu,\nu,\zeta)$
\cite{PER84},
where $\mu,\nu,\zeta$ are the complex variables corresponding
to the operators ${\hat a}_{1},{\hat b},{\hat a}_{2}$ respectively.
Realistic initial conditions 
are given by the vacuum state for the sideband modes, and a thermal state
at temperature $T$,
with mean vibrational number $\overline{n}=[\exp(\hbar\Omega/k_{B}T)-1]^{-1}$
for the mechanical mode ($k_{B}$ being the Boltzmann constant). 
Then, the state of the three-mode system is Gaussian, with characteristic
function
\begin{eqnarray}\label{eq:Phisol}
    \Phi&=&\exp\left[
    -{\cal A}|\mu|^{2}-{\cal B}|\nu|^{2}-{\cal E}|\zeta|^{2}
    +{\cal C}\mu\nu+{\cal C}\mu^{*}\nu^{*}\right.
    \nonumber\\
    &&+\left.{\cal F}\mu\zeta+{\cal F}\mu^{*}\zeta^{*}
    +{\cal D}\nu\zeta^{*}+{\cal D}\nu^{*}\zeta\right]\,,
\end{eqnarray}
where the coefficients ${\cal A}$,
${\cal B}$, ${\cal C}$, ${\cal D}$, ${\cal E}$, ${\cal F}$
depend on the interaction time $t$
\cite{LONG}, which is determined
by the time duration of the driving laser pulse.
The idea is now to find an experimentally feasible, {\em 
modified} version of the standard protocol for the teleportation 
of continuous quantum variables \cite{VAI94,BRA98}, able to minimize 
the disturbing effects of the beam-splitter-type term in Eq.~(\ref{eq:Heff}).
After reflection on the mirror, the sideband
modes ${\hat a}_{1}$ and ${\hat a}_{2}$
reach Alice's station.
Then Alice performs a heterodyne measurement \cite{YUE80}
on the mode ${\hat a}_{2}$, projecting it onto a coherent
state of complex amplitude $\alpha$.
Alice and Bob are left with an entangled state
for the optical Stokes mode $a_{1}$ and the vibrational mode $b$,
conditioned to this measurement result. 
In this case, the conditioned entangled state
is still Gaussian, and characterized by the following
correlation matrix (independent of $\alpha$, affecting only first 
order moments)
\begin{widetext}
\begin{equation}
    \Gamma =
    \left(
    \begin{array}{cccc}
    {\cal A}+\frac{1}{2} -\frac{{\cal F}^{2}}{{\cal E}+1}&0
    &{\cal C}+\frac{\cal FD}{{\cal E}+1}&0
    \\
    0&{\cal A}+\frac{1}{2}-\frac{{\cal F}^{2}}{{\cal E}+1}
    &0&-{\cal C}-\frac{\cal FD}{{\cal E}+1}
    \\
    {\cal C}+\frac{\cal FD}{{\cal E}+1}&0
    &{\cal B}+\frac{1}{2}-\frac{{\cal D}^{2}}{{\cal E}+1}&0
    \\
    0&-{\cal C}-\frac{\cal FD}{{\cal E}+1}&
    0&{\cal B}+\frac{1}{2}-\frac{{\cal D}^{2}}{{\cal E}+1}
    \end{array}
    \right)\,,
    \label{eq:Gam}
\end{equation}
\end{widetext}
where $\Gamma_{i,j}=\langle{\hat{\bf v}}_{i}{\hat{\bf v}}_{j}
+{\hat{\bf v}}_{j}{\hat{\bf v}}_{i}\rangle/2
-\langle{\hat{\bf v}}_{i}\rangle\langle{\hat{\bf v}}_{j}\rangle$ 
with ${\hat{\bf v}}=({\hat X}_{a_{1}},{\hat P}_{a_{1}},
{\hat X}_{b},{\hat P}_{b})$
and ${\hat X}_{a_{1}}=({\hat a}_{1}+{\hat a}_{1}^{\dag})/\sqrt{2}$,
${\hat P}_{a_{1}}=-i({\hat a}_{1}-{\hat a}_{1}^{\dag})/\sqrt{2}$,
${\hat X}_{b}=({\hat b}+{\hat b}^{\dag})/\sqrt{2}$,
${\hat P}_{b}=-i({\hat b}-{\hat b}^{\dag})/\sqrt{2}$.

Once Alice has performed the heterodyne measurement on 
${\hat a}_{2}$, one then follows
the standard protocol for 
continuous variables quantum teleportation \cite{VAI94,BRA98}.
The quantum channel between Alice and Bob is established
via the two-mode entangled state described by the correlation matrix
(\ref{eq:Gam}). Alice mixes the radiation mode provided by Victor, 
whose unknown state she wants to
teleport, with her part of entangled state (the Stokes $a_{1}$ mode),
on a balanced beam splitter
(see Fig.\ref{fig1}).
She then carries out a homodyne detection at each output port, thereby
measuring two commuting quadratures
${\hat X}_{+}=({\hat X}_{in}+{\hat X}_{a_{1}})/\sqrt{2}$ and
${\hat P}_{-}=({\hat P}_{in}-{\hat P}_{a_{1}})/\sqrt{2}$, with 
measurement results $ X_{+}$ and $P_{-}$.
The final step at the sending station is to transmit the classical
information, corresponding to the result of the heterodyne 
and homodyne measurements she
performed, to the receiving terminal.
Upon receiving this information, Bob displaces his part of entangled
state (the mirror acoustic mode) as follows:
${\hat X}_{b}\to {\hat X}_{b}+\sqrt{2}X_{+}
+\sqrt{2}{\rm Re}\{\alpha\}({\cal F}-{\cal D})/({\cal E}+1)$,
${\hat P}_{b}\to {\hat P}_{b}-\sqrt{2}P_{-}
+\sqrt{2}{\rm Im}\{\alpha\}({\cal F}+{\cal D})/({\cal E}+1)$.
Notice that, in our protocol, Bob's local operation depends
on all Alice's measurements ($X_{+}$, $P_{-}$, $\alpha$).
To actuate the phase-space displacement, 
Bob can use again the radiation pressure force.
In fact, if the mirror is shined by a bichromatic intense laser
field with frequencies $\varpi_{0}$ and
$\varpi_{0}+\Omega$, employing again Eq.~(\ref{eq:Hini})
and the RWA, one is left with an effective interaction Hamiltonian
$H_{act} \propto {\hat b}e^{-i\varphi}+{\hat b}^{\dag}e^{i\varphi}$,
where $\varphi$ is the relative phase between the two frequency 
components. Any phase space displacement of the mirror vibrational 
mode can be realized by adjusting this relative phase and the 
intensity of the laser beam.

\begin{figure}
\includegraphics[width=3.2in]{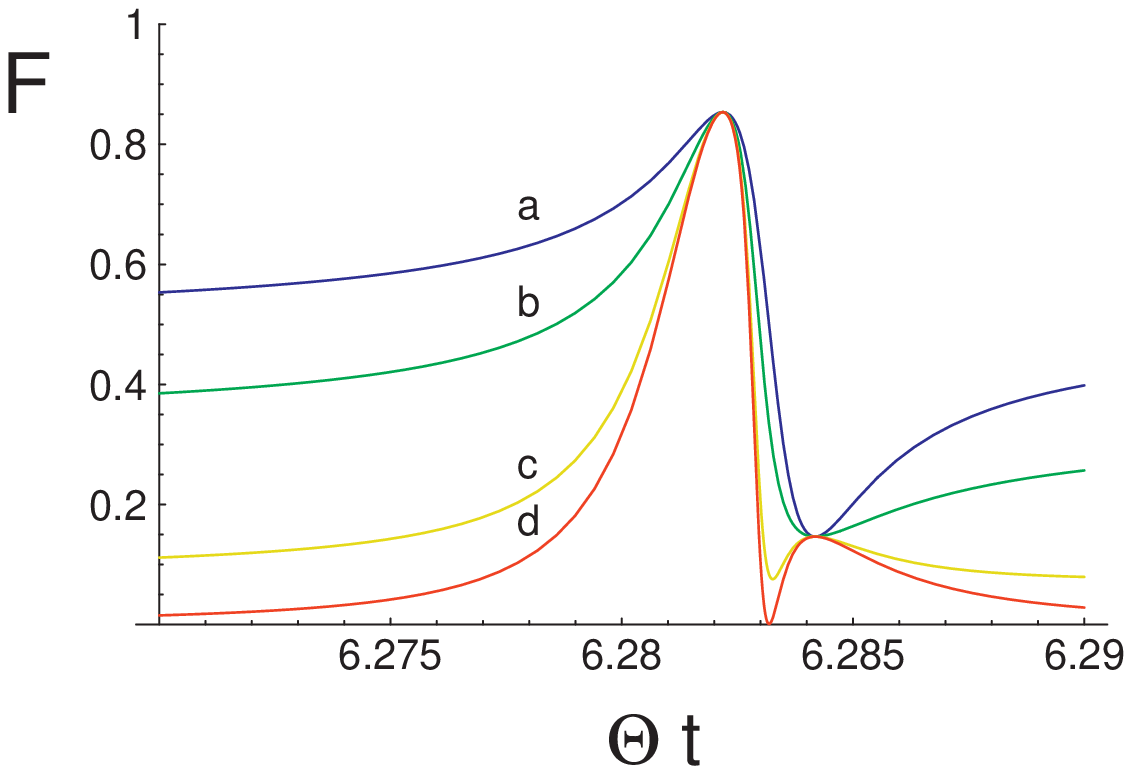}
\caption{
Fidelity $F$ vs the scaled time $\Theta t$, being 
$\Theta=\sqrt{\theta^{2}-\chi^{2}}$.
Curves a, b, c, d are for $\overline{n}=0$,
1, 10, $10^{3}$, respectively.
The values of parameters are: $\wp=10$ W;
$\Omega=5\times 10^{8}$ Hz;
$\Delta\nu_{det}=10^{7}$ Hz; $M=10^{-10}$ Kg;
$\omega_{0}= 2\times 10^{15}$ Hz, $\Delta\nu_{mode}=10^{3}$ Hz.}
\label{fig2}
\end{figure}

In the case when Victor provides an input Gaussian state 
characterized by a $2\times 2$ symmetric covariance matrix $\Gamma^{in}$,
the output state at Bob site is again Gaussian, with
a covariance matrix $\Gamma^{out}$.
The input-output relation for these matrices is given by \cite{CHI02}
\begin{eqnarray}
    \Gamma_{11}^{out}&=&\Gamma_{11}^{in}+\left(
    \Gamma_{11}+2\Gamma_{13}+\Gamma_{33}\right)\,,
    \label{eq:Gout1} \\
    \Gamma_{12}^{out}&=&\Gamma_{12}^{in}+\left(
    \Gamma_{14}-\Gamma_{12}+\Gamma_{34}-\Gamma_{23}\right)\,,
    \\
    \Gamma_{22}^{out}&=&\Gamma_{22}^{in}+\left(
    \Gamma_{22}-2\Gamma_{24}+\Gamma_{44}\right)\,. \label{eq:Gout3}
\end{eqnarray}
The fidelity $F$ of the described teleportation protocol,
defined as the overlap between the input and the output
state, can be written, with the
help of Eqs.~(\ref{eq:Gout1})-(\ref{eq:Gout3}) and (\ref{eq:Gam}), as
\begin{equation}
    F=\{1+\left[1+{\cal A}+{\cal B}+2{\cal C}
    -({\cal F}-{\cal D})^{2}/({\cal E}+1\right)]\}^{-1}
    \label{eq:F}
\end{equation}
where we have specialized to the case of an input coherent state.
In such a case, the upper bound for the fidelity achievable with only 
classical means and no quantum resources
is $F=1/2$ \cite{BRA99}. Fig.~\ref{fig2}
shows the fidelity as a function of the (rescaled) interaction time
$t$ for different values of the initial mean thermal phonon number of 
the mirror acoustic mode $\overline{n}$.
The fidelity $F$ is periodic
in the interaction time $t$,
and we show only one of all possible time windows where $F$ reaches
its maximum. The remarkable result shown in Fig.~\ref{fig2} 
is that this maximum value, 
$F_{max} \simeq 0.85$, is well above the classical bound $F=0.5$ and that it 
is surprisingly independent of the initial temperature of the acoustic mode.
This is apparently in contrast with previous results \cite{DUA00}
showing that entanglement is no longer useful above
one thermal photon (or phonon).
This effect could be ascribed to quantum interference
phenomena, and opens the way 
for the demonstration of quantum teleportation of states
of macroscopic systems.
However, thermal noise has still important effects so that, in 
practice, any experimental implementation needs an 
acoustic mode cooled at low temperatures (see however 
Refs.~\cite{COH99,VIT02}
for effective cooling mechanism of acoustic modes).
In fact, we see from Fig.~\ref{fig2} that by increasing $\overline{n}$,
the useful time interval becomes narrower.
That means the necessity of designing precise driving laser pulses
in order to have a well defined interaction time.
Furthermore, the time interval within
which the classical communication from Alice to Bob, and the phase 
space displacement by Bob have to be made, becomes shorter and shorter
with increasing temperature, because the vibrational state projected
by Alice's Bell measurement heats up in a time of the order 
of $(\gamma_{m}\overline{n})^{-1}$,
where $\gamma_{m}$ is the mechanical damping constant. The effects of 
mechanical damping can be instead neglected during the 
back-scattering process stimulated by the intense laser beam. In fact, 
mechanical damping rates of about $\gamma_{m} \simeq 1$ Hz are
available, and therefore negligible with respect to
the typical values of the coupling constants 
$\chi \simeq \theta \simeq 5\times 10^{5}$ Hz, and 
$\Theta=\sqrt{\theta^{2}-\chi^{2}}\simeq 
10^{3}$ Hz, determining the Hamiltonian dynamics.
Such values are obtained with the following choice of parameters:
${\wp} = 10$W, $\omega_{0}\sim 2\times 10^{15}$ Hz, $\Omega \sim 
5\times 10^{8}$ Hz, $\Delta\nu_{det}\sim 10^{7}$ Hz, 
$\Delta\nu_{mode}\sim t^{-1} \sim 10^{3}$ Hz, and $M \sim 10^{-10}$ 
Kg, which are those used in Fig.~\ref{fig2}. These parameters
are slightly different from those of already 
performed optomechanical experiments \cite{COH99,TIT99}.
However, using a thinner silica crystal and considering higher
frequency modes, the parameters we choose could be obtained. These choices
show the difficulties one meets in trying to extend
genuine quantum effects as teleportation into the macroscopic domain.
{\bf So far we have mainly focused on the
effects of the thermal noise acting on 
the mechanical resonator. Other fundamental noise sources are
shot noise and radiation pressure noise. 
Shot noise affects the detection of the sideband modes but its effect 
has been taken into account (it is responsible for the $1/2$ terms in the 
diagonal entries of the correlation matrix of Eq.~(\ref{eq:Gam})).
The radiation pressure noise due to the intensity fluctuations of the 
incident laser beam yields fluctuations of the optomechanical 
coupling. However, the chosen parameter values
give $\Delta\wp/\wp = \Delta (\Theta t)/2\Theta t \simeq 
10^{-8}$, which is negligibly small with respect to the width of the
window where the fidelity attains its maximum (see Fig.~2).}

The continuous variable teleportation protocol presented here modifies 
the standard one of Refs.~\cite{VAI94,BRA98} by adding a heterodyne 
measurement on the ``spectator'' mode ${\hat a}_{2}$. This additional 
measurement performed by Alice is important because it significantly 
improves the teleportation protocol. In fact, it is easy to see that 
if no measurement is performed on the anti-Stokes mode, 
the resulting fidelity for the teleportation of coherent states 
is always smaller with respect to that with the heterodyne measurement.
In particular, there is still a maximum value of the fidelity,
$F_{max}=0.80$ in this case, independent of temperature, 
but the useful interaction time interval becomes much narrower for 
increasing temperature. 

It is worth remarking that the present teleportation scheme provides
also a very powerful {\em cooling} mechanism for the acoustic mode
(see also \cite{COH99,VIT02}). 
As matter of fact, the effective number of thermal excitations of the
mirror state conditioned to the
homodyne measurements at Alice station becomes
$\overline{n}_{eff}=1+{\cal A}+{\cal B}+2{\cal C}
-({\cal F}-{\cal D})^{2}/({\cal E}+1)$.
It reduces to $\overline{n}+1$ in absence of entanglement,
where $1$ represents the noise introduced by the protocol.
Instead, the optomechanical interaction for a proper time permits
to achieve $\overline{n}_{eff}=0.17$, {\it i.e.}, an $80\%$ 
reduction of thermal noise.

Finally, for what concerns the experimental verification of
teleportation, that is, the measurement of the final state of the 
acoustic mode,
one can consider a second, intense ``reading'' laser pulse,
and exploit again the optomechanical interaction given by
Eq.~(\ref{eq:Heff}), where now $a_{1}$ and $a_{2}$ are meter modes.
It is in fact possible to 
perfom a heterodyne measurement of the combined sidebands mode
${\hat a}_{1}(t)-{\hat a}_{2}^{\dag}(t)$,
where $t$ is the time duration of the second laser pulse.
Solving the Heisenberg equations one obtains
\begin{eqnarray}
    &&{\hat a}_{1}(t)-{\hat a}_{2}^{\dag}(t)=
    \frac{1}{\Theta}\left[\chi+\theta\right]\sin(\Theta t) 
    {\hat b}^{\dag}(0)
    \nonumber\\
    &&+\frac{1}{\Theta^{2}}\left[
    \theta^{2}-\chi^{2}\cos(\Theta t)-\chi\theta+\chi\theta\cos(\Theta 
    t)\right]{\hat a}_{1}(0)
    \nonumber\\
    &&-\frac{1}{\Theta^{2}}\left[
    \chi\theta+\chi\theta\cos(\Theta t)-\chi^{2}-\theta^{2}\cos(\Theta 
    t)\right]{\hat a}_{2}^{\dag}(0)\,.
\end{eqnarray}
Then, it is easy to see that for $\cos\left(\Theta t\right)=0$
and $\Theta(\theta+\chi)\gg\theta(\theta-\chi)$
(as it is for the parameters values employed in Fig.~\ref{fig2}),
the measured quantity practically coincides with the mode
operator of the acoustic mode, 
thus revealing information on its state
(see also \cite{BRI02}).

In conclusion, we have proposed a simple scheme to teleport an
unknown quantum state of a radiation field
onto a macroscopic, collective vibrational 
degree of freedom of a massive mirror.
The basic resource of
entanglement is attained by means of the optomechanical coupling 
provided by the radiation pressure. 
Here we have shown the 
teleportation of the quantum information contained in an unknown 
quantum state of a radiation field to a collective degree of freedom 
of a massive object. This scheme could be easily extended in principle 
to realize a transfer of quantum information between two massive 
objects. In fact
Victor could use tomographic reconstruction schemes, 
again based on the ponderomotive interaction (see \cite{MAN96}), to 
``read'' the quantum state of a vibrational mode of another mirror
and use this information to prepare the state of the radiation field 
to be sent to Alice.
The present result could be challenging tested
with present technology, and opens new perspectives towards the use
of quantum mechanics in the macroscopic world.
For example, we recognize possible
technological applications such as the
preparation of nonclassical states of
MEMS \cite{CLE98},
where the oscillation frequency could be higher and, consequently, the
working temperature can be raised.


\begin{thebibliography}{99}
\bibitem{LONG}
The detailed calculations will be presented elsewhere 
for space reasons.
\end{thebibliography}
\end{document}